%% file: main.tex
\documentclass{article}
\usepackage[utf8]{inputenc}
\usepackage{amsmath}
\usepackage{hyperref}
\usepackage{graphicx}
\usepackage{braket}
\graphicspath{ {./Images/} }

\input{NewCommands}

\begin{document}

\begin{center}
{\Large\textbf{\mathversion{bold} Notes on a non-thermal fluctuation-dissipation relation in quantum Brownian motion}
\par}

\vspace{0.8cm}

\textrm{Xinyi Chen-Lin\footnote{xinyitsenlin@gmail.com}}
\vspace{4mm}

\textit{NORDITA\\
Hannes Alfvéns väg 12,
SE-106 91 Stockholm, Sweden}

\vspace{5mm}

\textbf{Abstract} 
\vspace{5mm}

\begin{minipage}{13cm}
\input{Sections/abstract}

\end{minipage}

\end{center}

\newpage

% \maketitle

\tableofcontents

\input{Sections/Intro}

\input{Sections/openQM}

\input{Sections/QBM}
\input{Sections/conclusion}

\input{Sections/acknowledgement}
\appendix
\input{Appendix/SK}

\input{Appendix/FDT}

\input{Appendix/GLE}

\input{Appendix/notationConvention}

\bibliographystyle{ourbst}
\bibliography{refs}

\end{document}

%% file: NewCommands.tex
\newcommand{\noise}{F}

\newcommand{\tr}{\mathrm{Tr}\,}
\newcommand{\ft}[1]{\tilde{#1}}

\newcommand{\intR}{\int_{-\infty}^{\infty}}
\newcommand{\figref}[1]{Fig.\,\ref{#1}}
\DeclareMathOperator{\csch}{csch}

%% file: Sections/abstract.tex
We review how unitarity and stationarity in the Schwinger-Keldysh formalism naturally lead to a (quantum) generalized fluctuation-dissipation relation (gFDR) that works beyond thermal equilibrium. Non-Gaussian loop corrections are also presented. Additionally, we illustrate the application of this gFDR in various scenarios related to quantum Brownian motion and the generalized Langevin equation.

%% file: Sections/Intro.tex
\section{Introduction}

The Schwinger-Keldysh (SK) formalism \cite{schwinger1961brownian, keldysh1965diagram} is a well-known path integral framework to study quantum non-equilibrium many-body systems \cite{kamenev2023field}. In the classical limit, it reduces to the Martin-Siggia-Rose (MSR) path integral, which is also equivalent to the stochastic Langevin description. 

One important aspect of the SK approach is that symmetries become a powerful tool to derive important relations in both quantum and classical regimes. For example, the thermodynamic equilibrium can be formulated as a symmetry of the SK action \cite{Sieberer_2015}, and leads to the celebrated fluctuation-dissipation theorem (FDT). Another example is the breaking of the time-reversal symmetry giving rise to fluctuation theorems in non-equilibrium physics \cite{aron2010symmetries, aron2018non}.
Besides, the path integral approach allows a systematic and perturbative way to deal with non-linear theories beyond the scope of the standard Langevin approach.

These notes focus on the unitarity constraint on the SK two-point correlation functions in the stationary limit, which can be viewed as a generalized fluctuation-dissipation relation (gFDR) that works beyond thermal stationary states and Markov approximation. While the classical limit of this gFDR is certainly known in the study of aging glassy systems, where it is used to define an effective temperature \cite{Cugliandolo_2011}, its use in the broader classical stochastic community does not seem widespread, in contrast to the use of the non-Markovian generalization of the Langevin equation (GLE).

The objective of these notes is not to provide an extensive review of existing literature\footnote{The present work was conducted mostly independently and with limited awareness of the literature on glassy systems. It is part of my journey to learn about classical and quantum non-equilibrium frameworks.}. Instead, the aim is to present a self-contained derivation of the gFDR for non-driven non-equilibrium systems and illustrate its application through various examples.

The paper is structured as follows. In section \ref{sec:openQ}, we review the SK framework for open quantum systems with one effective degree of freedom. Then, we impose the stationary condition at the generating functional level. Next we perturbatively derive the gFDR in Fourier space. In section \ref{sec:QBM}, the gFDR is applied to GLE and quantum Brownian motion (QBM). Finally, we conclude in section \ref{sec:conclusion}. In the appendix, we provide additional review material on the SK formalism in section \ref{sec:SK}, FDT in section \ref{sec:FDT}, and GLE in section \ref{sec:Langevin}.

%% file: Sections/openQM.tex
\section{Effective SK theories}\label{sec:openQ}

\subsection{Open quantum systems}
Consider a quantum system whose action is separable:
\begin{equation}
    S=S_\text{subsystem} + S_\text{environment} + S_\text{interaction}.
\end{equation}
By integrating out the environment degrees of freedom, we obtain an SK effective theory for the subsystem, which is equivalent to the Feynman-Vernon theory \cite{feynman1963theory}.
The SK generating functional, see \ref{sec:generatingFunctional}, can be rewritten in terms of the reduced density matrix $\rho_r$ and the reduced time evolution kernel $J_r$:
\begin{align}
    Z(t,t_0; \phi,\phi']&=\intR dx_0\intR dx_0' \rho_r(x_0,x_0',t_0) \intR dx \, J_r(x,x,t, x_0,x_0',t_0; \phi,\phi'] 
\end{align}
with
\begin{align}
    J_r(x,x',t,x_0,x_0', t_0 ; \phi,\phi']
    &=\int_{q(t_0)=x_0}^{q(t)=x} \mathcal{D} q \int_{q'(t_0)=x_0'}^{q'(t)=x'} \mathcal{D} q' \, \exp{\left(\frac{i}{\hbar} A_\text{CG}[q,q'; \phi,\phi']\right)},
\end{align}
where the subindex CG emphasizes a coarse-grained effective action, and $\phi, \phi'$ are the external sources. The curly/square parenthesis notation refers to function/functional of the left/right parameters inside, separated by a semi-colon.
This action includes the action that affects only the subsystem, and the influence phase $\Phi[q,q']$, which summarizes all the contributions from the environment, its initial state and its interaction with the subsystem:
\begin{align}
    A_\text{CG}[q,q'; \phi,\phi'] &= S_\text{subsystem}[q]-S_\text{subsystem}[q']+\Phi[q,q'] \nonumber\\
    &+\int_{t_0}^{t} ds\, (q(s)\phi(s)- q'(s)\phi'(s)).
\end{align}
Here, it is also assumed that the initial density matrix factorizes as a product of the reduced density matrix and that of the environment. It does not matter, though, as we are interested only in the stationary regime.
% where the environmental 'noise' must erase all the information from the initial distribution. 

\subsection{Stationary regime} \label{sec:stationarity}

Let us define the \emph{stationary regime} as the one with the correlators being
\begin{enumerate} 
    \item independent on the initial conditions and distribution $\rho(x_0,x_0',t_0)$,
    \item time-translation invariant.
\end{enumerate}

\input{Sections/GaussianPI}

\input{Sections/quadratic}

%% file: Sections/GaussianPI.tex
\subsection{Generating functional for stationary states}
Consider a quadratic effective action $A_0[q,q'; \phi, \phi']$.
Its corresponding time evolution kernel is a double Gaussian path integral. Therefore, it is solvable and splits into the classical path contribution and fluctuations around the classical path:
\begin{align}
    J_r(x,x,t, x_0,x_0',t_0;\phi,\phi'] 
    &= \int_{q(t_0)=x_0}^{q(t)=x} \mathcal{D} q \int_{q'(t_0)=x_0'}^{q'(t)=x} \mathcal{D} q' \, \exp{\left(\frac{i}{\hbar} A_0[q,q'; \phi,\phi']\right)}\\
    &= C(t,t_0)   \exp{\left(\frac{i}{\hbar} S_\mathrm{cl}(x,x_0,x_0'; \phi,\phi']\right)},
\end{align}
where the integration over quantum fluctuations hidden in $C(t,t_0)$, does not depend on the external sources.

The classical action is separable into a term independent of the initial conditions and a term that is not:
\begin{align}
 S_\mathrm{cl}(x,x_0,x_0; \phi,\phi']&= 
    S_\mathrm{cl}(x,0,0; \phi,\phi']
    +\delta S(x,x_0,x_0',\phi,\phi']\\
    &\approx S_\mathrm{cl}^\infty(x,0,0; \phi,\phi'], \quad (t\longrightarrow \infty)
\end{align}
where we imposed the condition of independency on the initial distribution, in the stationary limit. The time-translational invariance condition allows us to take the initial and final times to be minus and plus infinities. Finally, the generating functional for stationary states is simplified to:
\begin{align}
    Z_{0}^{\infty}[\phi,\phi'] &\equiv Z_0(\infty,-\infty;\phi,\phi']\\
    &=C(\infty,-\infty) \intR dx \,   \exp{\left(\frac{i}{\hbar} S_\mathrm{cl}^\infty(x,0,0; \phi,\phi']\right)},
\end{align}
since the integrations over the initial distribution is just one. 
% In other words, the action evaluated at the classical path encodes all the steady state information.

For quadratic theories perturbed by an external potential $V(q,q')$, see appendix \ref{sec:perturbation}, we can do perturbation theory as usual:
\begin{align}\label{eq:ZVstationary}
    Z_{V}^{\infty}[\phi,\phi']&=\exp\left(- \frac{i}{\hbar}\int_{-\infty}^{\infty} dt\, V\left(\frac{\hbar}{i}\frac{\delta}{\delta \phi}, \frac{\hbar}{i}\frac{\delta}{\delta \phi'}\right)\right)Z_{0}^{\infty}[\phi,\phi'].
\end{align}

%% file: Sections/quadratic.tex
\subsection{Quadratic effective action}
In the stationary limit, the most general SK quadratic effective action (the external currents are set to zero here for simplicity) in Keldysh basis is as follows:
\begin{align}\label{action}
    A_0[q_r, q_a]
    =&\int_{-\infty}^{\infty} dt' \, q_a(t')  \left\{(D_{ar}* q_r)(t')
    + i   (D_{aa}* q_a)(t') \right\},
\end{align}
where the star operation is the Fourier convolution. Note that $q_r-q_r$ term is prohibited by unitarity, see appendix \ref{sec:SK}. 
The kernel $D_{ar}$ contains time derivative operators, and $D_{ar}$ is non-zero and satisfies
\begin{equation}
    D_{aa}(-t)=D_{aa}(t).
\end{equation}
% such that
% \begin{equation}
%     \intR dt' \, x_a(t') (D_{aa}* x_a)(t') = \frac{1}{2} \intR dt' \intR dt'' \, x_a(t') D_{aa}(t'-t'') x_a(t'').
% \end{equation}

The action above can be identified with the stochastic Martin-Siggia-Rose action, where $q_a(t)$ plays the role of the auxiliary field (see also appendix of A of \cite{hsiang2018quantum}). Therefore,
it is also equivalent to a generalized stationary Langevin equation with a Gaussian noise $F(t)$:
\begin{align}
    -(D_{ar}*q_r)(t)&=\noise(t)\\
    \frac{1}{2}\braket{\{\noise(t), \noise(t')\}}&=D_{aa}(t-t').
\end{align}
where the curly parenthesis represents the anticommutator.

\subsection{Green's function}
\begin{figure}
    \centering
    \includegraphics[scale=0.35]{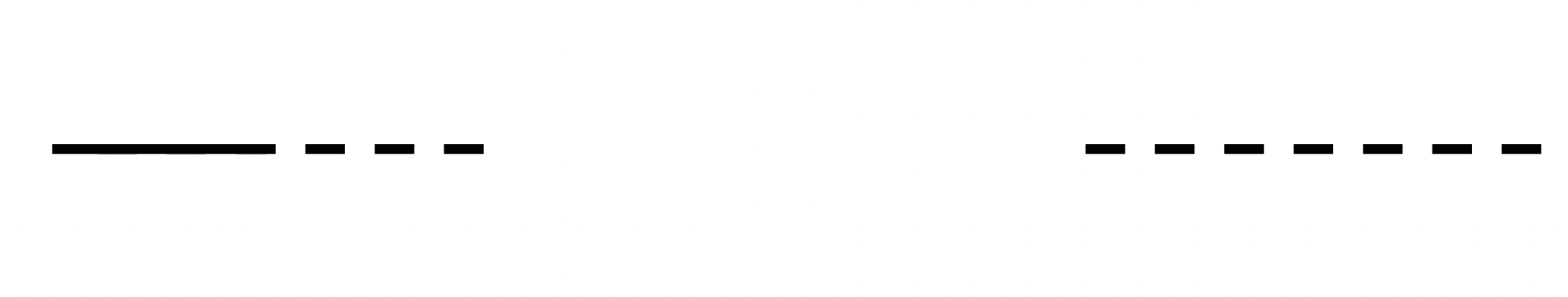}
    \caption{Feynman diagram for the retarded Green's function $G_{ra}$ on the left and the noise kernel $D_{aa}$ on the right. The solid line represents $q_r$ and the dashed line is for $q_a$. }
    \label{fig:feynDiagrams}
\end{figure}

Because the action \eqref{action} is time-translation invariant, we can Fourier transform it:
\begin{align}
    \ft{A}_0[\ft{q}_r, \ft{q}_a]
    =&\frac{1}{2\pi}\int_{-\infty}^{\infty} d\omega \, \ft{q}_a(-\omega)  \left\{\ft{D}_{ar}(\omega)  \ft{q}_r(\omega)
    + i   \ft{D}_{aa}(\omega)  \ft{q}_a(\omega)  \right\},
\end{align}
from which two Feynman diagrams can be read off, see Fig. \ref{fig:feynDiagrams}. The $q_a-q_a$ diagram is the interaction vertex, and the $q_r-q_a$ diagram is the retarded Green's function, which is:
\begin{equation}
\ft{G}^R(\omega)=-\frac{1}{\ft{D}_{ar}(\omega)},
\end{equation}
since it is the solution of the homogeneous equation:
\begin{equation}
-(D_{ar}*G^R)(t)=\delta(t).
\end{equation}
Of course, the retarded Green's function must satisfy the causality condition
\begin{align}\label{eq:causalityG}
    G^R(t<0)=0,
\end{align}
which implies that the lower half plane in the frequency $\omega$-space must be analytic.

\begin{figure}
    \centering
    \includegraphics[scale=0.35]{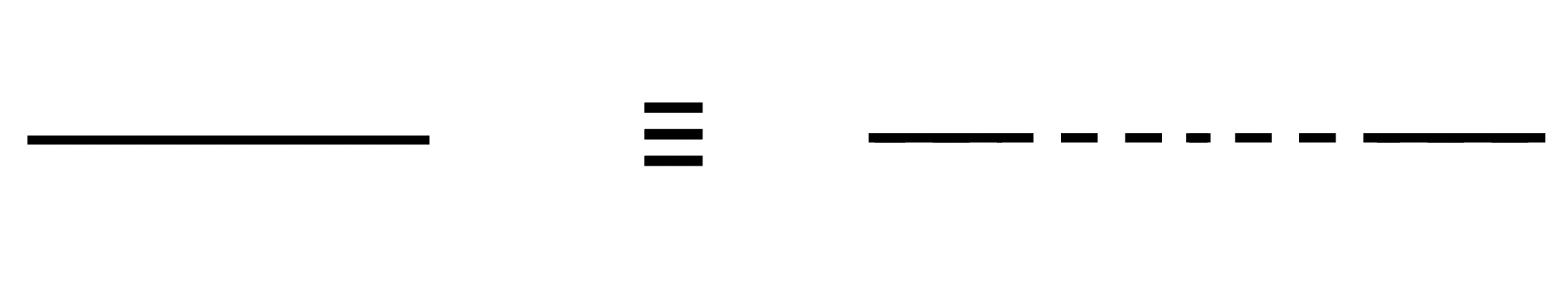}
    \caption{Feynman diagram for the symmetric Green's function  $G_{rr}$.}
    \label{fig:GssFeynDiagram}
\end{figure}

The autocorrelator, or the symmetric Green's function \eqref{defsGs}, is a simple composite diagram (see Fig. \ref{fig:GssFeynDiagram}) in this limit:
\begin{equation}\label{eq:GS_composition}
    \ft{G}^S(\omega)
    =  \left|\ft{G}^{R}(\omega) \right|^2 \ft{D}_{aa}(\omega).
    % =\left|\ft{G}^R(\omega) \right|^2 \ft{D}_{aa}(\omega)
\end{equation}
For complex retarded Green functions, we use the following identity 
\begin{align}
   \mathrm{Im}\ft{G}^R(\omega) = \left|\ft{G}^{R}(\omega) \right|^2 \mathrm{Im} \ft{D}_{ar}(\omega) ,
\end{align}
to obtain:
\begin{equation}\label{GsImGr}
   \boxed{ \frac{\ft{G}^S(\omega)}{\mathrm{Im}\ft{G}^R(\omega)}
    = \frac{ \ft{D}_{aa}(\omega)}{\mathrm{Im} \ft{D}_{ar}(\omega)}} .
\end{equation}
Given that $\ft{G}^S(\omega)$ is even in $\omega$ space, the inverse Fourier transform can be written as:
\begin{align}\label{GsImGr-Time}
    G^S(t)
    &= \frac{1}{\pi}\int_0^\infty d\omega \, \ft{G}^S(\omega) \cos(\omega t)\\
    & = \frac{1}{\pi}\int_0^\infty d\omega \frac{ \ft{D}_{aa}(\omega)}{\mathrm{Im} \ft{D}_{ar}(\omega)} \mathrm{Im}\ft{G}^R(\omega-i 0^+) \cos(\omega t), \quad (t\longrightarrow \infty)
\end{align}
where the latter expression (we added the prescription for the poles of the retarded Green's function) is valid only if the stationarity conditions in section \ref{sec:stationarity} are fulfilled.

We will see some examples, but before that, let us generalize the above relation for non-quadratic theories.

\input{Sections/loop}

%% file: Sections/loop.tex
\subsection{Loop corrections} \label{sec:loop}
\begin{figure}
    \centering
    \includegraphics[scale=0.25]{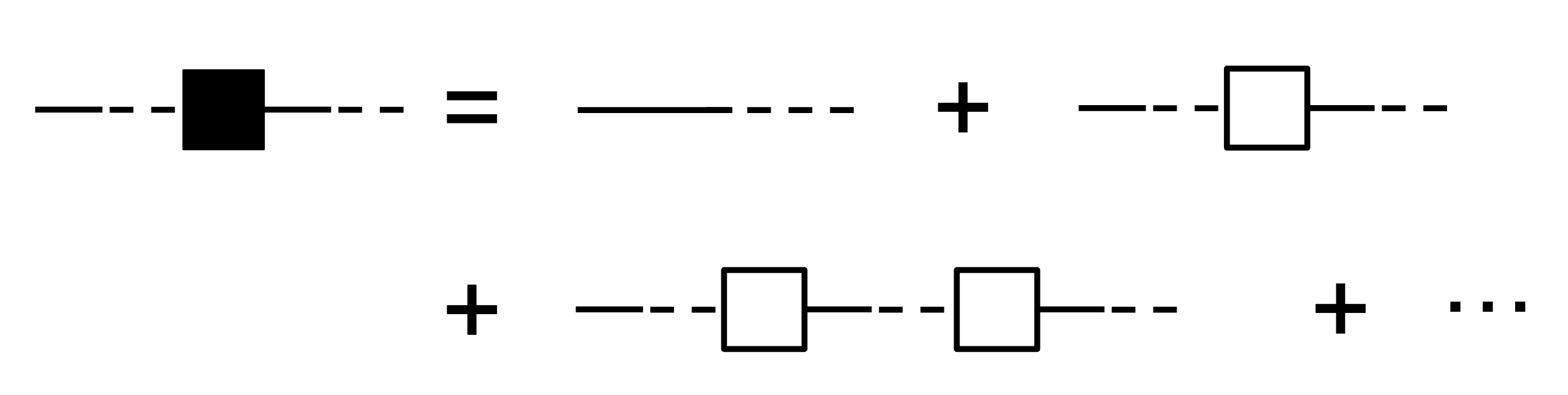}
    \caption{Renormalized retarded Green's function. The self-energy $\delta\ft{D}_{ar}(\omega)$ is the white square. }
    \label{fig:Gr_corrected}
\end{figure}

Small non-quadratic terms added to the action can be treated perturbatively according to \eqref{eq:ZVstationary}. For the retarded Green's function, also known as the propagator, the loop corrections (aka self-energy) which we call $\delta \ft{D}_{ar}(\omega)$, form a geometric sum as shown in \figref{fig:Gr_corrected}: 
\begin{align}
    \ft{G}_r^R(\omega)
    &=\ft{G}^R(\omega) (1 +  \delta \ft{D}_{ar}(\omega) \,\ft{G}^R(\omega)  + \delta \ft{D}_{ar}(\omega)^2\, \ft{G}^R(\omega)^2+ \ldots ) \\
    &= \frac{\ft{G}^R(\omega)}{1-\delta \ft{D}_{ar}(\omega)\, \ft{G}^R(\omega)}.
\end{align}
The equation above is known as the Dyson equation. In terms of the kernel $\ft{D}_{ar}(\omega)$, which is the negative inverse of the propagator, the above expression simplifies to: 
\begin{align}
     \ft{D}_{ar,r}(\omega)&= \ft{D}_{ar}(\omega)+\delta \ft{D}_{ar}(\omega).
\end{align}
Note that we use an additional subindex $r$ to stand for renormalized/corrected values, and it should not be confused with the $r$ from the Keldysh variables.

\begin{figure}
    \centering
    \includegraphics[scale=0.35]{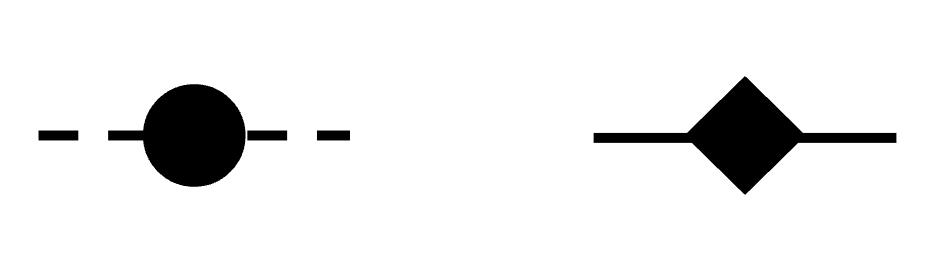}
    \caption{Corrections to the interaction vertices. The solid line represents $q_r$ and the dashed line $q_a$. The black circle is $\delta \ft{D}_{aa}(\omega)$ and the diamond is $\delta Q(\omega)$.}
    \label{fig:vertices}
\end{figure}

The correction to the other kernel, shown as the black circle in \figref{fig:vertices}, is simply additive:
\begin{equation}
    \ft{D}_{aa,r}(\omega) = \ft{D}_{aa}(\omega) + \delta \ft{D}_{aa}(\omega).
\end{equation}

\begin{figure}
    \centering
    \includegraphics[scale=0.22]{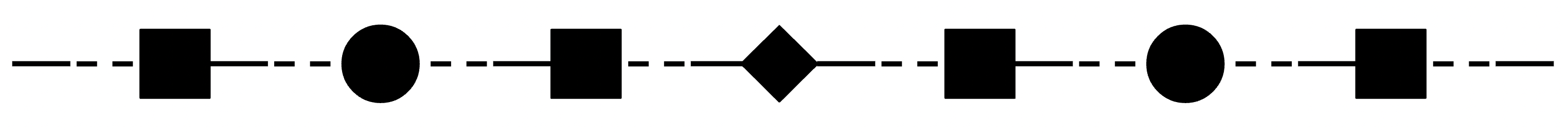}
    \caption{The renormalized symmetric Green's function is composed of the renormalized retarded Green's function, see \figref{fig:Gr_corrected}, and the vertex corrections in \figref{fig:vertices}.}
    \label{fig:GsCorrections}
\end{figure}

\begin{figure}
    \centering
    \includegraphics[scale=0.22]{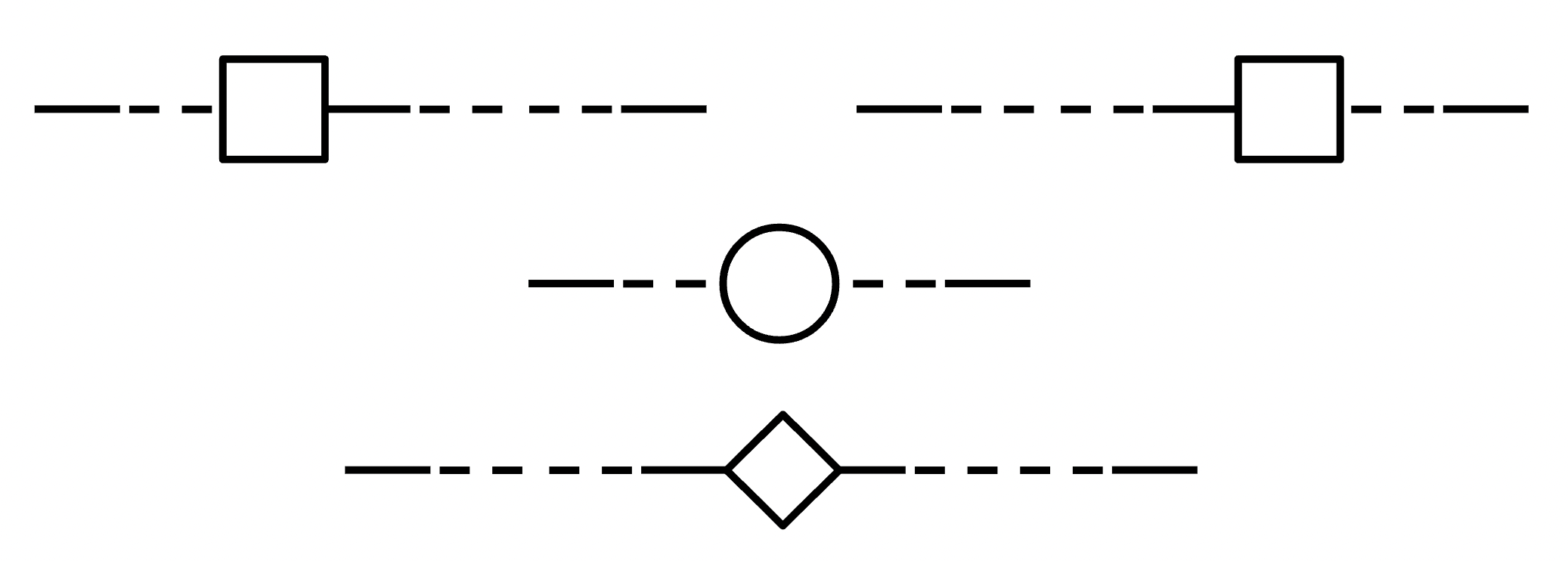}
    \caption{The 1-loop contributions (the white geometric shapes are symbolic only) to the renormalized symmetric Green's function.}
    \label{fig:Gs-1loop}
\end{figure}

Now, new interaction vertices connecting $q_r$ with $q_r$ can emerge, and we call it $\delta Q(\omega)$ and represent it as the diamond in \figref{fig:vertices}. This type of interaction breaks the tree level 
composition rule \eqref{GsImGr}. Nevertheless, we can still write down a generic loop-corrected expression for the symmetric Green's function in a closed form, which corresponds to \figref{fig:GsCorrections}:
\begin{align}
    \ft{G}_r^S(\omega) = \ft{D}_{aa,r}(\omega) \left|\ft{G}_r^R(\omega)\right|^2 (1+ \delta Q(\omega) \ft{D}_{aa,r}(\omega) \left|\ft{G}_r^R(\omega)\right|^2  ).
\end{align}
Replacing the modulus squared by the imaginary part:
\begin{align}
   \mathrm{Im}\ft{G}^R_r(\omega) =  \left|\ft{G}_r^R(\omega)\right|^2 
   \mathrm{Im} \ft{D}_{ar,r}(\omega) ,
\end{align}
we can rewrite the above expression as:
\begin{align}
    \frac{\ft{G}_r^S(\omega)}{ \mathrm{Im}\ft{G}^R_r(\omega)} &= \frac{\ft{D}_{aa,r}(\omega)}{\mathrm{Im} \ft{D}_{ar,r}(\omega)} 
    \left( 1+\delta  Q(\omega) \frac{\ft{D}_{aa,r}(\omega)}{\mathrm{Im} \ft{D}_{ar,r}(\omega)}  \mathrm{Im}\ft{G}^R_r(\omega)  \right). \label{eq:loopFDT}
\end{align}
This expression is useful when we Taylor-expand the right-hand-side (RHS) in terms of the unperturbed quantities and the loop contributions of different orders. For the leading order correction, we get
\begin{equation}\label{eq:1loopFDT}\boxed{
    \frac{\ft{G}_r^S(\omega)}{ \mathrm{Im}\ft{G}^R_r(\omega)}
     \approx  \frac{\ft{G}^S(\omega)}{ \mathrm{Im}\ft{G}^R(\omega)}
    \left( 1+ \frac{\delta \ft{D}_{aa}(\omega)}{\ft{D}_{aa}(\omega)}-\frac{\mathrm{Im} \delta \ft{D}_{ar}(\omega)}{\mathrm{Im}\ft{D}_{ar}(\omega)}+\delta Q(\omega) \,\ft{G}^S(\omega)  \right)},
\end{equation}
where the three corrections are shown in \figref{fig:Gs-1loop}.

\subsection{Thermalization condition}\label{sec:thermalization}
If thermalization happens, both \eqref{GsImGr} and \eqref{eq:loopFDT} must satisfy FDT \eqref{quantumFDT}:
\begin{align}
    \frac{\ft{G}^S(\omega)}{ \mathrm{Im}\ft{G}^R(\omega)} &= \hbar \coth \frac{\hbar \beta \omega}{2},\\
    \frac{\ft{G}_r^S(\omega)}{ \mathrm{Im}\ft{G}^R_r(\omega)} &= \hbar \coth \frac{\hbar \beta_r \omega}{2}.
\end{align}
Therefore
\begin{align}
    \coth \frac{\hbar \beta_r \omega}{2}
    & \approx
    \coth \frac{\hbar \beta \omega}{2}
    \left( 1+ \Delta(\omega) \right) \label{eq:qFDTresult}
\end{align}
where we use $\Delta(\omega)$ as a short-hand notation for the loop contributions to the ratio of the Green's functions, for example, at the leading order, see \eqref{eq:1loopFDT}, it is:
\begin{equation}\label{eq:Delta}
    \Delta(\omega)= 
    \frac{\delta \ft{D}_{aa}(\omega)}{\ft{D}_{aa}(\omega)}-\frac{\mathrm{Im} \delta \ft{D}_{ar}(\omega)}{\mathrm{Im}\ft{D}_{ar}(\omega)}+\delta Q(\omega) \,\ft{G}^S(\omega).
\end{equation}
Note that $\Delta(\omega)$ can depend on the temperature, but its dependence is not made explicit here.

On the other hand, FDT is a non-perturbative result. Therefore, we must systematically expand the hyperbolic cotangent in terms of the corrections to the inverse temperature, namely $\beta_r=\beta+\delta\beta$, as $\delta\beta$ shall absorb the loop contributions in the thermal case. Then, the expansion is to be compared order by order with the one from \eqref{eq:qFDTresult}. Let us show how it works for the leading order. First:
\begin{align}
    \coth \frac{\hbar \beta_r \omega}{2}
    & =
    \coth \frac{\hbar (\beta+\delta\beta) \omega}{2}\\
    &\approx \coth \frac{\hbar \beta \omega}{2}
    \left(1-\hbar \delta\beta \omega \csch \hbar\beta \omega \right). \label{eq:cothExpand}
\end{align}
Then, comparing the RHS of \eqref{eq:qFDTresult} with \eqref{eq:cothExpand}, we see that thermalization condition at this order implies:
\begin{equation} \label{eq:thermalization}
    \delta\beta\overset{!}{=}
    -\frac{\Delta(\omega)}
    {\hbar  \omega \csch\hbar\beta \omega}.  
\end{equation}
Since $\delta\beta$ must be a constant, there are three possible scenarios for thermalization:
\begin{align} \label{eq:scenarios}
    \Delta(\omega) \begin{cases}
      \propto \omega \csch\hbar\beta \omega & \text{Quantum thermalization}\\
      = 0 & \text{Quantum thermalization: } \beta_r=\beta\\
      = \text{constant} & \text{Classical/high-T  thermalization}\\  
    \end{cases}    
\end{align}
Again, this a perturbative result, hence the conclusion for (quantum or classical) thermalization is only valid up to the level studied. This means higher order terms could potentially drive the system away from thermalization. On the other hand, the negative statement about thermalization is sufficient with only perturbative information.

%% file: Sections/QBM.tex
\section{Quantum Brownian motion}\label{sec:QBM}
In this section, we focus on one of the simplest integrable SK effective theories, that is, a quantum Brownian motion. This is a quantum harmonic oscillator coupled linearly to a harmonic bath. The classical action is the sum of the following actions:
\begin{align}
S[x]=&  \int^t_{0} ds \, \frac{1}{2} M\left(\dot{x}(s)^{2}-\Omega_R^{2} x(s)^{2}\right) \\
S_\mathrm{bath}[\{x_n\}]=&\int^t_{0} ds
\sum_{n} \frac{1}{2} m_{n}\left(\dot{x}_{n}(s)^{2}-\omega_{n}^{2}x_{n}(s)^2\right)
\\
S_\mathrm{int}[x, \{x_n\}]=&-\int^t_{0} ds
\sum_{n}  c_{n}x_{n}(s) x(s)
\end{align}
where 
\begin{equation}
    \Omega_R^{2}=\Omega^{2}+\delta\Omega^{2}, \quad \delta \Omega^2 =  \sum_{n}  \frac{c_{n}^2}{2 m_n \omega_n^2} .
\end{equation}

In the continuum limit\footnote{The finite version is also integrable, but thermalization happens only in the continuum limit.}, the bath is fully characterized by a spectral density function $I(\omega)$, that is related to the discrete frequencies as follow:
\begin{equation}\label{spectralDensity}
    I(\omega)=\sum_{n} \frac{c_{n}^{2}}{2 m_{n} \omega_{n}} \delta\left(\omega-\omega_{n}\right).
\end{equation}
Traditionally, power law distributions $I(\omega)\propto \omega^\alpha$ were studied, where $\alpha=1$ is called the Ohmic bath, and when $\alpha$ is smaller or bigger than 1, sub-Ohmic or supra-Ohmic baths.

The Feynman-Vernon/SK approach gives rise to an effective quadratic action \eqref{action} with the kernels \cite{caldeira1983path,  hu1992quantum}:
\begin{align}
    D_{ar}(t) &= -M \delta(t) (\partial^2_t + \Omega_R^2) -  \mu(t)\\
    D_{aa}(t) &= \nu(t)
\end{align}
where $\mu$ and $\nu$ are known as the dissipation and the noise kernel, respectively. In this case of a linearly coupled bath, the dissipation kernel in Fourier space is fully determined by the spectral density of the bath, \cite{ford1988quantum, caldeira2014introduction}\footnote{Note that \cite{caldeira2014introduction} absorbs the $\pi$ in the definition of the spectral density.}:
\begin{equation}\label{eq:muI}
    \ft{\mu}(\omega)= i \pi I(\omega).
\end{equation}

This problem is solvable for all time  \cite{fleming2011exact}, and it is equivalent to the stationary generalized Langevin equation, see appendix \ref{sec:Langevin}. 
 The Fourier transform of the retarded Green's function is:
\begin{equation}\label{eq:GR_QBM}
   \ft{G}^R(\omega)
   =\frac{1/M}{-\omega^{2}+\Omega_R^{2}+\ft{\mu}(\omega)/M},
\end{equation} 
and the symmetric Green's function in the stationary limit is obtained from \eqref{GsImGr}:
\begin{equation} \label{eq:GsImGr_QBM}
    \frac{\ft{G}^S(\omega)}{\mathrm{Im} \ft{G}^R(\omega)}=\frac{\ft{\nu}(\omega)}{ \mathrm{Im}\ft{\mu}(\omega) }.
\end{equation}

\subsection{Bath in thermal equilibrium}
If the bath is in initial thermal equilibrium, the noise kernel is:
\begin{align}
    \nu(t) &=\int_{0}^{\infty} d \omega\, \hbar
    \coth\left(\frac{\hbar\beta \omega}{2} \right) I(\omega)  \cos \omega t,
\end{align}
which in Fourier space is just the FDT for the bath (sometimes referred as 2FDT in the classical literature):
\begin{equation}\label{eq:bathFDT}
 \frac{\ft{\nu}(\omega) }{\mathrm{Im} \ft{\mu}(\omega)}= \hbar \coth\left(\frac{\hbar \beta  \omega}{2} \right)  \xrightarrow{\hbar \rightarrow 0 } \frac{2}{\beta \omega}.
\end{equation}
If the bath were to be modeled by a scalar field, $\nu$ and $\mu$ would correspond precisely to the symmetric and retarded Green's functions of the field, see \cite{hsiang2018quantum}. 

Now, the gFDR \eqref{eq:GsImGr_QBM} implies:
\begin{equation}
    \frac{\ft{G}^S(\omega)}{\mathrm{Im} \ft{G}^R(\omega)}=\hbar \coth\left(\frac{\hbar \beta \omega}{2} \right) \xrightarrow{\hbar \rightarrow 0 } \frac{2}{\beta \omega}.
\end{equation}
In other words, when the system eventually thermalizes, the final temperature is the same as the one of the initial bath. This case is also studied and emphasized in \cite{hsiang2018quantum}. The FDT above is sometimes called 1FDT in the classical literature, and it will differ from 2FDT in non-equilibrium scenarios.

\subsection{Multiple baths}
The sum of Gaussian random variables are Gaussian, hence, when there are many harmonic baths, the gFDR \eqref{eq:GsImGr_QBM}is easily generalized to:
\begin{equation}
    \frac{\ft{G}^S(\omega)}{\mathrm{Im} \ft{G}^R(\omega)} =\frac{\sum_i\ft{\nu}_i(\omega)}{ \sum_i\mathrm{Im}\ft{\mu}_i(\omega) }.
\end{equation}
Even when the baths are in thermal equilibrium (i.e. satisfying \eqref{eq:bathFDT}), it is clear from the above expression that, in general, the system does not thermalize. It was also shown in \cite{aron2010symmetries} that the multi-bath setting breaks the thermal symmetry. Only in the classical limit, when all the baths have the same spectral density, then, the particle thermalizes at an average temperature of the baths:
\begin{equation}
    \frac{\ft{G}^S(\omega)}{\mathrm{Im} \ft{G}^R(\omega)} =\frac{2 k_B \bar{T}}{\omega }  .
\end{equation}
This case has been numerically studied for realistic thermal baths in \cite{ness2016nonequilibrium}. 

The above expression can be generalized to effective temperatures for classical out-of-equilibrium baths, see \cite{zamponi2005fluctuation}.

\subsection{Out-of-equilibrium bath}
The generalized Langevin equation is used to model classical glassy systems with a slow relaxation dynamics \cite{cugliandolo1994off}. The gFDR \eqref{eq:GsImGr_QBM} naturally appears in this context, where an effective temperature \cite{Cugliandolo_2011} defined for the bath encodes the non-equilibrium properties.

In this section, we apply the gFDR to a relatively recent example of a semiclassical time glass model in \cite{verstraten2021time}, that was solved using fractional calculus. The microscopic model is characterized by the following noise kernel and the spectral density (hence the dissipation kernel through \eqref{eq:muI})\footnote{The paper \cite{verstraten2021time} defines their spectral density $J(\omega)$ that differs a factor $\pi$ from our definition, i.e. $J(\omega)= \pi I(\omega)$.}:
\begin{align}
    \nu (t) &=2\int _0^{\infty } d\omega \frac{  (t_s \omega)^{1-s}}{\beta  \omega } I(\omega) \cos (\omega t)\\
    I(\omega) &= \frac{\eta}{\pi}  \sin\left(\frac{\pi}{2} s\right) \omega^s,
\end{align}
where $0<s<1$. The integration gives:
\begin{align}
    \nu (t) 
    &=\frac{2}{\pi } \frac{ t_s^{1-s}}{\beta}  \eta  \sin\left(\frac{\pi}{2} s\right)\int _0^{\infty } d\omega  \cos (\omega t)\\
    &= \frac{2 t_s^{1-s}}{\beta}  \eta  \sin\left(\frac{\pi}{2} s\right)\delta(t).
    % &=\frac{2 \pi\, t_s^{1-s} }{\beta \omega^s}I(\omega)\delta(t).
\end{align}
The prefactor before the Dirac delta function is the Fourier transform of the noise kernel, therefore, together with \eqref{eq:muI}, the relation \eqref{eq:GsImGr_QBM} reduces to:
\begin{align} \label{eq:GsImGr_TLS}
     \frac{\ft{G}^S(\omega)}{\mathrm{Im} \ft{G}^R(\omega)}=\frac{2 t_s^{1-s} }{ \beta \omega^s}.
\end{align}
The effective frequency-dependent inverse temperature is then:
\begin{equation}
    \beta_\text{eff}(\omega) = (\omega t_s)^{s-1} \beta.
\end{equation}
When $s=1$, which corresponds to an Ohmic bath, we recover the classical FDT \eqref{eq:classicalFDT}. According to \cite{verstraten2021time}, only when $s<1$, i.e. in the sub-Ohmic regime, the non-equilibrium time glassy behavior appears.

\subsection{Anomalous diffusion}
The diffusion of a classical Brownian particle is characterized by the late-time behavior of the mean square displacement (MSD):
\begin{equation}
    MSD(t)=\left\langle[q(t)-q(0)]^{2}\right\rangle.
\end{equation}
Ford and O'Connell in \cite{ford2006anomalous} related MSD to FDT and studied anomalous diffusion for non-Ohmic baths. Their relation for the time derivative of MSD can be readily generalized to:
\begin{equation}\label{eq:MSD'}
    % MSD'(t)
    % =  \frac{1}{\pi}\int_0^\infty d\omega \frac{ \ft{\nu}(\omega)}{\mathrm{Im} \ft{\mu}(\omega)} \mathrm{Im}\ft{G}^R(\omega-i 0^+)  \sin(\omega t).
    MSD'(t) = \int_0^\infty d\omega\, \ft{G}^S(\omega) \omega \sin(\omega t)
\end{equation}
where the symmetric Green's function can be determined by the gFDR \eqref{eq:GsImGr_QBM}.

Let us apply this method to the non-equilibrium time glass model in the subsection above, and we will reproduce the asymptotic results in the supplemental material of \cite{verstraten2021time}. Recall \eqref{eq:GsImGr_TLS} and the retarded Green's function \eqref{eq:GR_QBM} (with free particle as done in \cite{verstraten2021time}):
\begin{align}
   \ft{G}_S(\omega)
    % &=\frac{2 t_s^{1-s} }{\beta \omega^s}\mathrm{Im} \ft{G}^R(\omega)\\
    &=\frac{2 t_s^{1-s} }{\beta \omega^s}\frac{\pi I(\omega) M^{-2}}{\omega^4+ \pi^2 I(\omega)^2 M^{-2}}\\
    &\propto  \frac{1}{\omega^4+ A^2 \omega^{2s}}\\
    &\propto  \frac{1}{\omega^{2s}(\omega^{4-2s}+ A)}\\
    &\propto \frac{1}{\omega^{2s}}, \quad (\omega\approx 0)
\end{align}
where $A=M^{-1} \eta \sin\left(\frac{\pi}{2} s\right)$, and we took the small frequency limit in order to obtain the late-time behavior for the time derivative of the MSD \eqref{eq:MSD'}, which is
\begin{align}
    MSD'(t) & \propto  \int_0^\infty d\omega\,  \frac{\sin(\omega t)}{\omega^{2s-1}}\propto t^{2s-2}
    % MSD'(t) & \propto  \int_0^\infty d\omega\,  \sin(\omega t) \propto \frac{1}{t}, \quad s= 1/2\\
    % MSD'(t) & \propto  \int_0^\infty d\omega\,  \frac{\sin(\omega t)}{\omega} \propto \mathrm{constant}, \quad s=1
\end{align}
hence, for MSD, we recover the following asymptotic results:
\begin{align}
    MSD(t)&\propto \log(t), \quad s=1/2,\\
    MSD(t)&\propto  t^{2s-1}, \quad s\neq 1/2.
\end{align}

% \subsection{Non-linear coupling }
% When the coupling to the bath is non-linear, but weak, \cite{hu1993quantum} used perturbation theory to find non-linear corrections to the dissipation and memory kernels, for $k=2,3,4$. We will not reproduce their results here, but will comment briefly that, despite assuming the bath to be in initial thermal equilibrium, plugging in their results in \eqref{eq:GsImGr_QBM} shows that the qFDT \eqref{quantumFDT} is not fulfilled, hence no quantum thermalization, except in the high and low temperature limit, as they discussed in the section III of their paper \cite{hu1993quantum}.  

\subsection{Anharmonic oscillator}\label{sec:anharmonic}
Now, let us show an example for the loop-corrected gFDR \eqref{eq:1loopFDT}.

\begin{figure}
    \centering
    \includegraphics[scale=0.25]{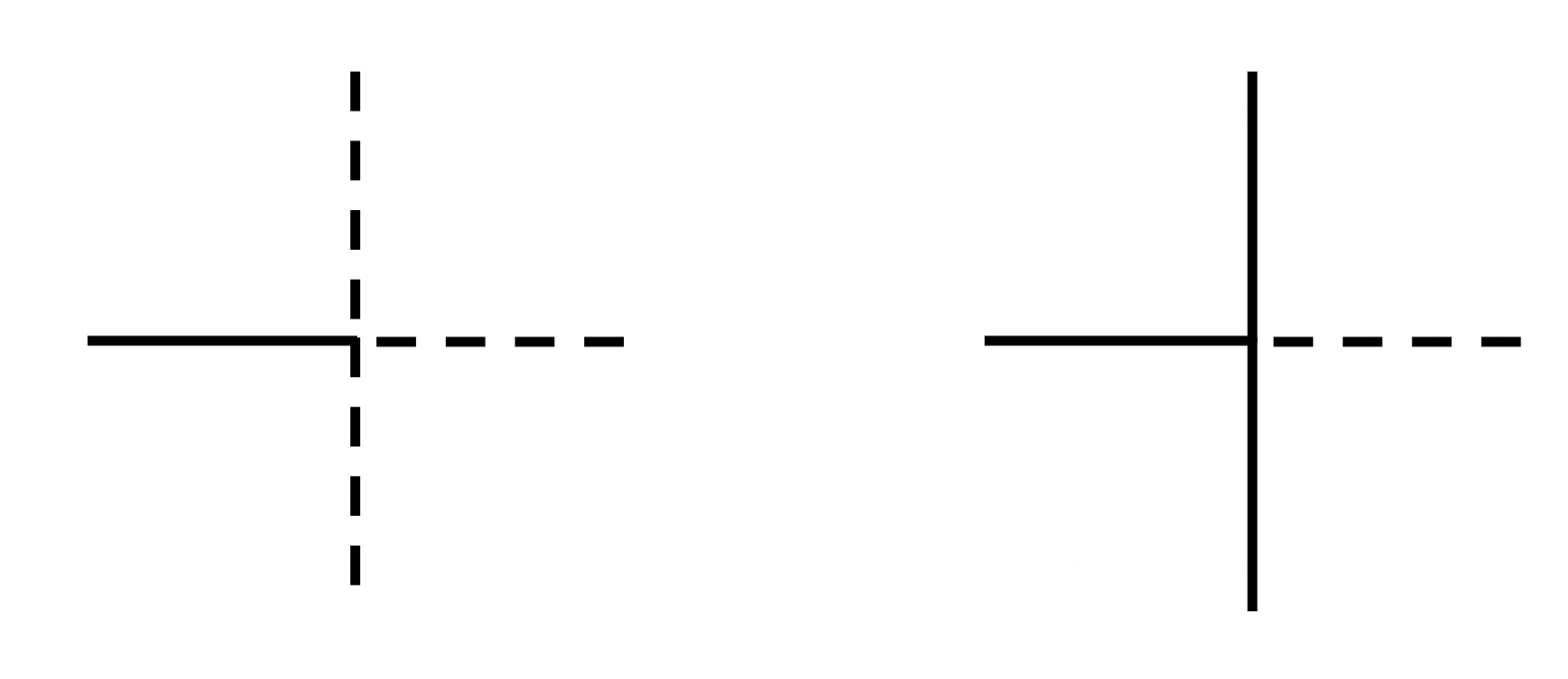}
    \caption{The quartic interaction vertices}
    \label{fig:quarticVertices}
\end{figure}

\begin{figure}
    \centering
    \includegraphics[scale=0.25]{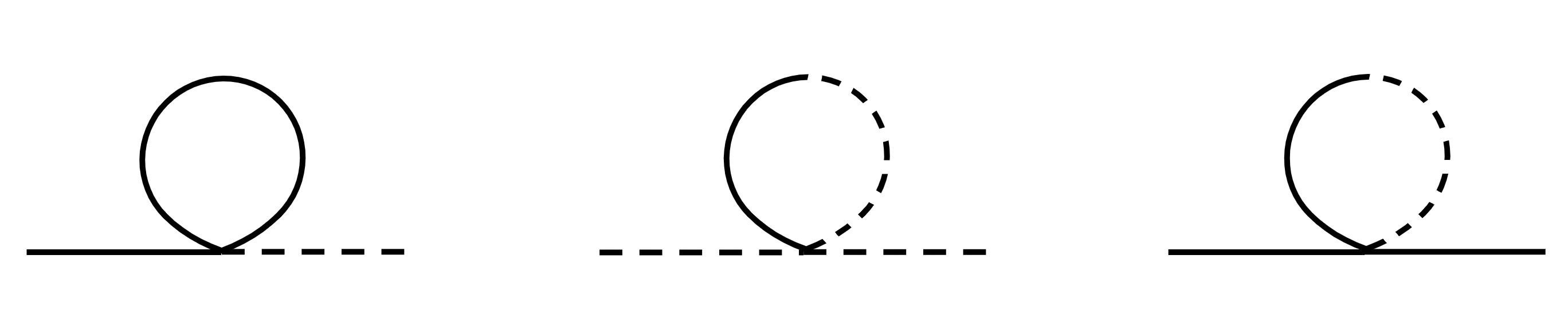}
    \caption{Leading order corrections. However, the diagrams with retarded Green's function loop (the two right ones) are vanishing because of causality. Their corresponding complex conjugate ones are not shown.}
    \label{fig:quartic1}
\end{figure}

\begin{figure}
    \centering
    \includegraphics[scale=0.25]{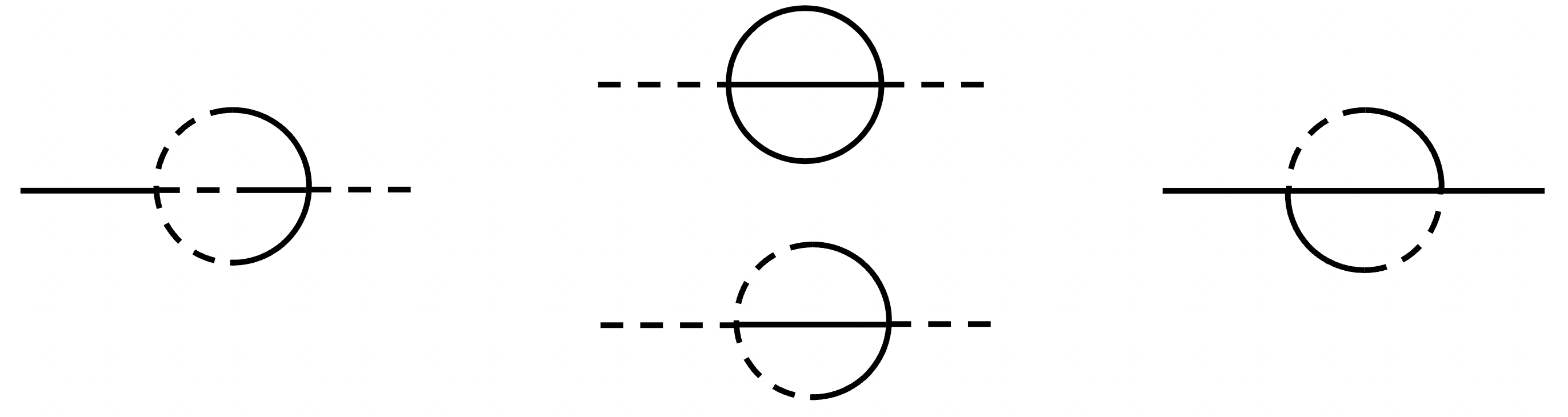}
    \caption{Next-to-leading order corrections. Their corresponding complex conjugate ones are not shown.}
    \label{fig:quartic2}
\end{figure}

Instead of the harmonic oscillator, let us consider a quartic anharmonic oscillator subject to the following SK potential:
\begin{align}
    V(q,q') =\frac{\lambda}{4!} (q^4 - q'^4)
\end{align}
which in Keldysh basis becomes
\begin{align}
    V(q_r,q_a)=\frac{\lambda}{4!} (q_r q_a^3 - 3 q_r^3 q_a),
\end{align}
where the two contributions are diagrammatically represented in \figref{fig:quarticVertices}. With these diagrams, we can build the leading order loop corrections shown in \figref{fig:quartic1}, which correspond to the $\delta \ft{D}_{ar}$, $\delta \ft{D}_{aa}$ and $\delta Q$ term in \eqref{eq:loopFDT}, respectively. However, the loop integral with the retarded Green's function vanishes because of causality. Furthermore, the loop integral with the symmetric Green's function is real. That means, at leading order in $\lambda$, the qFDT is \emph{not} corrected. Hence, at the leading order perturbation, the anharmonic oscillator thermalizes with the same temperature as the harmonic oscillator, as discussed in subsection \ref{sec:thermalization}. The loop computation was done explicitly by Hsiang et al \cite{hsiang2020nonequilibrium}, and that is precisely their result. Furthermore, the same authors have a non-perturbative argument for thermalization discussed in \cite{Hsiang:2020fbz}. Then, it would be interesting to compute the next-to-leading order corrections, where the Feynman diagrams are shown in \figref{fig:quartic2}, because these do not seem to vanish and could potentially violate the thermalization condition.

% \subsection{Non-linear coupling to the bath}
% When the bath of harmonic oscillators is not linearly coupled to the quantum particle, but weakly, perturbation theory was used in \cite{hu1993quantum} to derive an effective quadratic theory. 

% We will not reproduce their results here, but comment briefly that, despite assuming the bath to be in initial thermal equilibrium, plugging
% in their results in (40) shows that the qFDT (110) is not fulfilled, hence no
% quantum thermalization, except in the high and low temperature limit, as they
% discussed in the section III of their paper [13]
% \todo{Expand this section?}

%% file: Sections/conclusion.tex
\section{Conclusion}\label{sec:conclusion}

In summary, we presented the generalized fluctuation-dissipation relation (gFDR) \eqref{GsImGr} for the quadratic Schwinger-Keldysh (SK) path integral and its loop generalization \eqref{eq:loopFDT}. 
The gFDR generalizes the fluctuation-dissipation theorem (FDT) to non-thermal baths, providing valuable insight into FDT itself and its breakdown in various scenarios. We were also able to reproduce the asymptotic behavior of the correlation function and the anomalous diffusion for the time glass model in \cite{verstraten2021time}. Besides, the loop gFDR \eqref{eq:loopFDT} can be used to determine thermalization if the quadratic system is perturbed.

Beyond the presented cases, equations \eqref{GsImGr} and \eqref{eq:loopFDT} apply to other effective SK theories in the stationary limit, such as the SK effective field theory for diffusion \cite{Crossley:2015evo, Chen-Lin:2018kfl}. This theory, however, is thermal by construction, leading to the reduction of the gFDR to FDT for all loop orders, as one can check \cite{Jain_2021}. Another example of an SK effective theory where these equations apply is a non-linear SK diffusion model without an external driven force, which was proposed to model a quantum time crystal state in \cite{hayata2018diffusive}. 
Unitarity and stationarity can also be used to generalize the non-linear quantum FDTs presented in \cite{wang2002generalized} to non-linear gFDRs. Finally, for driven systems, there is some work done towards a FDR for non-equilibrium steady states in \cite{Hsiang:2020vus}.

%% file: Sections/acknowledgement.tex
\subsection*{Acknowledgements}
We would like to thank K.~Zarembo, S.~Krishnamurthy and E.~Aurell for discussion and reviewing the manuscript. We would also like to thank L.~Cugliandolo for pointing out some relevant existing work. 
This work was supported by the Knut and Alice Wallenberg Foundation.

%% file: Appendix/SK.tex
\section{Schwinger-Keldysh} \label{sec:SK}
In this section, we briefly review the Schwinger-Keldysh formalism in the coordinate respresentation. The reader can learn more from \cite{kamenev2023field, bentov2021schwinger, glorioso2018lectures}.

\subsection{Density operator}
Consider a time-evolving quantum statistical system characterized by the density operator:
\begin{align}
    \rho(t) 
    &=\frac{1}{N} \sum_{n=1}^{N} p_{n} \ket{\psi_{n}(t)} \bra{\psi_{n}(t) },
\end{align}
such that the statistical weights $p_n$ sum to one, and the evolution of the quantum states $\ket{\psi_n(t)}$ is governed by the Schroedinger equation:
\begin{equation}
    i \hbar \frac{d}{dt} \ket{\psi_n(t)} = H  \ket{\psi_{n}(t)},
\end{equation}
whose solution is
\begin{equation}
    \ket{\psi_n(t)} = U(t-t_0) \ket{\psi_n(t_0)}
\end{equation}
with the evolution operator being
\begin{equation}
    U(t)=e^{-i t H /\hbar}.
\end{equation}
Therefore the evolution of the density operator is:
\begin{equation}
\rho(t)=U\left(t-t_0\right) \rho(t_0) U^{\dagger}\left(t- t_0\right).
\end{equation}

\subsection{Time-evolution kernel}
The density operator in the coordinate basis is:
\begin{align}
    \rho(x,x',t)
    &=\frac{1}{N} \sum_{n=1}^N p_n \, \psi_n(x,t)\psi_n(x',t)^*\\
    &= \intR dx_0 \, \intR dx_0'\,  J(x,x',t,x_0,x_0',t_0 )
     \rho(x_0,x_0',t_0),
\end{align}
where the time-evolution kernel is a double copy of the quantum mechanical time-evolution kernel:
\begin{align}
    J(x,x',t,x_0,x_0', t_0 )
    &\equiv\langle x \vert U(t-t_0) \vert x_0\rangle 
    \langle x' \vert U(t-t_0) \vert x'_0\rangle^*\\
    &=\int_{q(t_0)=x_0}^{q(t)=x} \mathcal{D} q \int_{q'(t_0)=x_0'}^{q'(t)=x'} \mathcal{D} q' \, \exp{\left(\frac{i}{\hbar} A[q,q']\right)},
\end{align}
where the action is:
\begin{equation}
    A[q,q']=S[q]-S[q'].
\end{equation}
Notice that the time-evolution kernel is time-translation invariance, because of the unitary evolution.

% \subsection{Partition Function}
% The partition function is:
% \begin{align}
%     Z &= \tr \rho(t)\\
%       &= \intR dx \, \rho(x,x,t)\\
%       &= \,\intR dx_0 \, \intR dx_0'\,\rho(x_0,x_0',t_0) \intR dx\, J(x,x,t,x_0,x_0', t_0 )  \label{eq:partitionFunction}\\
%       &=1
% \end{align}
% which is normalized to one as shown in the last identity.

\subsection{Generating functional}\label{sec:generatingFunctional}
We can probe the system by minimally coupling it to an external source $\phi(t)$, that means: 
\begin{align}
    J(x,x,t,x_0,x_0', t_0; \phi,\phi']
    &=\int_{q(t_0)=x_0}^{q(t)=x} \mathcal{D} q \int_{q'(t_0)=x_0'}^{q'(t)=x'} \mathcal{D} q' \, \exp{\left(\frac{i}{\hbar} A[q,q'; \phi,\phi']\right)}\\
    A[q,q';\phi,\phi']&= S[q]-S[q']+\int_{t_0}^{t} ds\, (q(s)\phi(s)- q'(s)\phi'(s)). \label{eq:actionA}
\end{align}
Then the partition function of this system is the generating functional for the correlation functions of the original system:
\begin{align}\label{eq:generatingFunctional}
    Z[\phi,\phi']
      &= \,\intR dx_0 \, \intR dx_0'\,\rho(x_0,x_0',t_0) \intR dx\, J(x,x,t,x_0,x_0', t_0 ; \phi,\phi'].
\end{align}
In particular, its logarithm 
\begin{equation}
    W[\phi,\phi']\equiv\ln Z[\phi,\phi']
\end{equation}
generates connected correlators.

\subsubsection{Properties}
The unitary evolution of the density matrix implies the  following properties for the SK generating functional:
\begin{enumerate}
    \item Normalization condition: 
    \begin{equation}\label{normalization}
        Z[\phi,\phi]=1 \longrightarrow W[\phi, \phi']=0
    \end{equation}
    \item Reflection symmetry:
    \begin{equation}\label{reflection}
        Z[\phi',\phi]=Z[\phi,\phi']^*
    \longrightarrow W[\phi',\phi]=W[\phi,\phi']^*
    \end{equation}
\end{enumerate}

\subsection{Perturbation theory} \label{sec:perturbation}
For a quadratic action $A_{0}[q,q';\phi,\phi']$ perturbed with a non-quadratic potential $V(q, q')$:
\begin{align}
    A_V[q,q';\phi,\phi']= A_{0}[q,q';\phi,\phi']-\int_{t_0}^{t} ds\, V(q(s), q'(s)),
\end{align}
the corresponding generating functional of the resulting action $A_V[q,q';\phi,\phi']$ can be expressed in terms of the unperturbed one:
\begin{align}\label{eq:ZV}
    Z_V[\phi,\phi']&=\exp\left(- \frac{i}{\hbar}\int_{t_0}^{t} ds\, V\left(\frac{\hbar}{i}\frac{\delta}{\delta \phi}, \frac{\hbar}{i}\frac{\delta}{\delta \phi'}\right)\right)Z_0[\phi,\phi']\\
    &\approx 
    \left(1 - \frac{i}{\hbar}\int_{t_0}^{t} ds\, V\left(\frac{\hbar}{i}\frac{\delta}{\delta \phi}, \frac{\hbar}{i}\frac{\delta}{\delta \phi'}\right) + \ldots \right)Z_0[\phi,\phi'],
\end{align}
where in the last step, we expanded the exponential. 

\subsection{Keldysh basis}
In practice, it is more convenient to work in the so-called Keldysh basis, i.e.:
\begin{align}\label{eq:Keldysh_basis}
    % \phi_r&=\frac{1}{2} (\phi+\phi'), \quad \phi_a=\phi-\phi'\\
    q_r&=\frac{1}{2} (q+q'), \quad q_a=q-q',
\end{align}
that also applies to the external currents.
Then, the action \eqref{eq:actionA} becomes:
\begin{align}
    A[q_r,q_a;\phi_r,\phi_a]= S[q_r,q_a]+\int_{t_0}^{t} ds\, (q_r(s)\phi_a(s)+q_a(s)\phi_r(s)).
\end{align}
% and the generating functional:
% \begin{align}
%     Z_V[\phi_r,\phi_a]&=\exp\left(- \frac{i}{\hbar}\int_{t_0}^{t} dt\, V\left(\frac{\hbar}{i}\frac{\delta}{\delta \phi_a}, \frac{\hbar}{i}\frac{\delta}{\delta \phi_r}\right)\right)Z_0[\phi_r,\phi_a].
%     % &\approx 
%     % \left(1 - \frac{i}{\hbar}\int_{t_0}^{t} dt\, V\left(\frac{\hbar}{i}\frac{\delta}{\delta \phi_a}, \frac{\hbar}{i}\frac{\delta}{\delta \phi_r}\right) + \ldots \right)Z_0[\phi_r,\phi_a],
% \end{align}

\input{Appendix/GF}

%% file: Appendix/GF.tex
\subsection{Green's functions}
The two-point (path-ordered) correlation functions in Keldysh basis are defined below:
\begin{align}
    G_{rr}(t_1,t_2)&\equiv\braket{\mathcal{P}q_r(t_1) q_r(t_2)} 
    =\left(\frac{\hbar}{i}\right)^2\left.\frac{\delta^2 W[\phi_r,\phi_a]}{\delta \phi_a(t_1)\phi_a(t_2)}\right|_{\phi_{r,a}=0} 
    \\
    \frac{\hbar}{i} G_{ra}(t_1,t_2)&\equiv\braket{\mathcal{P}q_r(t_1) q_a(t_2)} 
    =\left(\frac{\hbar}{i}\right)^2\left.\frac{\delta^2 W[\phi_r,\phi_a]}{\delta \phi_a(t_1)\phi_r(t_2)}\right|_{\phi_{r,a}=0} 
    \\
    \frac{\hbar}{i} G_{ar}(t_1,t_2)&\equiv \braket{\mathcal{P}q_a(t_1) q_r(t_2)} 
    =\left(\frac{\hbar}{i}\right)^2\left.\frac{\delta^2 W[\phi_r,\phi_a]}{\delta \phi_r(t_1)\phi_a(t_2)}\right|_{\phi_{r,a}=0} 
    \\
    \left(\frac{\hbar}{i}\right)^2 G_{aa}(t_1,t_2)&\equiv\braket{\mathcal{P}q_a(t_1) q_a(t_2)} 
    =\left(\frac{\hbar}{i}\right)^2\left.\frac{\delta^2 W[\phi_r,\phi_a]}{\delta \phi_r(t_1)\phi_r(t_2)}\right|_{\phi_{r,a}=0}=0 
    .
\end{align}
The latter one vanishes due to the normalization condition \eqref{normalization}. In fact, \eqref{normalization} implies that all
\begin{equation}
    G_{a\ldots a}=0.
\end{equation}
This identity is sometimes called \emph{Schwinger-Keldysh collapse rule} in the literature. 
% In other words, there are no $q_r^n$ terms in the action, where $n$ is a positive integer.

The reflection symmetry \eqref{reflection} implies:
\begin{equation}
    G_{ra}(t_1,t_2)=G_{ar}(t_2,t_1),
\end{equation}
i.e. it relates the retarded and the advanced Green's functions, since the non-vanishing two-point correlators are identified with the symmetric, retarded and advanced Green's functions:
\begin{align}\label{defsGs}
    G_{r r}(t_1,t_2)&=G^S(t_1,t_2)\equiv\frac{1}{2}\langle\{q(t_1),q(t_2)\}\rangle\\
    G_{r a}(t_1,t_2)&=G^R(t_1,t_2)\equiv \frac{i}{\hbar}\, \theta(t_1-t_2)\Delta(t_1,t_2)\\
    G_{a r}(t_1,t_2)&=G^A(t_1,t_2)\equiv -\frac{i}{\hbar}\, \theta(t_2-t_1)\Delta(t_1,t_2),
\end{align}
where
\begin{equation}
    \Delta(t_1,t_2) 
    =\langle[q(t_1),q(t_2)]\rangle.
\end{equation}

Hence, for a quadratic theory, the influence phase is simply:
\begin{align}
    W_0[\phi_r, \phi_a] &= \frac{i}{ \hbar} \frac{1}{2} \int_{t_0}^{t} dt_1 \int_{t_0}^{t} dt_2\,
    \Phi(t_1)^T
    \mathbf{G}(t_1,t_2)
    \Phi(t_2)
\end{align}
where
\begin{align}
    \Phi
    &=\left(\begin{array}{l}
    \phi_r \\
    \phi_a
    \end{array}\right), \quad 
    \mathbf{G}
    % &=\left(\begin{array}{cc}
    % G_{a a} & G_{a r} \\
    % G_{r a} & \frac{i}{\hbar} G_{r r}
    % \end{array}\right)
    =\left(\begin{array}{cc}
    0 & G^A \\
    G^R & \frac{i}{\hbar} G^S
    \end{array}\right).
\end{align}

%% file: Appendix/FDT.tex
\section{Fluctuation-Dissipation Theorem} \label{sec:FDT}
When the density matrix is thermal, i.e. a Gibbs state:
\begin{align}
    \rho_{\beta}=\frac{1}{Z_{\beta}} e^{-\beta H}, \quad Z_{\beta}=\tr\left(e^{-\beta H}\right),
\end{align}
then the generating functional satisfies the so-called Kubo-Martin-Schwinger (KMS) condition \cite{kubo1957statistical,martin1959theory}. In the case of thermal two-point correlators, the KMS condition translates into the well-known quantum fluctuation-dissipation theorem (FDT). Let us show how to derive it. First,
\begin{align}
    \langle O_i(t_1) O_j(t_2)\rangle
    &= \frac{1}{Z_\beta}\tr [
    e^{-\beta H} U^\dagger(t_1-t_0) O_i U(t_1-t_2) O_j U(t_2-t_0)]\nonumber \\
    &= \frac{1}{Z_\beta}\tr [
    e^{-\beta H} U^\dagger(t_1-t_0) O_i  \nonumber \\ 
    &\qquad \qquad U(t_1-t_0)U(t_0-t_2) O_j U(t_2-t_0)] \nonumber\\
    &= \frac{1}{Z_\beta}\tr [
    e^{-\beta H} e^{\beta H}U(t_0-t_2) O_j U(t_2-t_0)e^{-\beta H} U^\dagger(t_1-t_0)  \nonumber \\ 
    &\qquad \qquad O_i  U(t_1-t_0)] \nonumber\\
    &= \frac{1}{Z_\beta}\tr [
    e^{-\beta H} U(t_0-t_2+i \hbar \beta) O_j U(t_2-i\hbar \beta -t_1)  O_i  U(t_1-t_0)] \nonumber\\
    &=\langle O_j(t_2-i\hbar \beta) O_i(t_1)\rangle. \label{eq:OO}
\end{align}
Now, applying the following operator identities:
\begin{align}
    A B &= \frac{1}{2}\{A,B\}+\frac{1}{2}[A,B]\\
    B A &= \frac{1}{2}\{A,B\}-\frac{1}{2}[A,B]
\end{align}
and the definitions of the Green's functions in \eqref{defsGs}, the identity \eqref{eq:OO} becomes
\begin{align}
    G^S_{i j}(t_1-t_2)+\frac{1}{2}\Delta_{i j}(t_1-t_2) 
    &= G^S_{i j}(t_1-t_2+i\hbar \beta)-\frac{1}{2}\Delta_{i j}(t_1-t_2+i\hbar \beta).
\end{align}
Its Fourier transform is simply:
% \begin{align}
%     \ft{G}^S_{i j}(\omega)+\frac{1}{2}\ft{\Delta}_{i j}(\omega) 
%     &= e^{\hbar \beta \omega}\ft{G}^S_{i j}(\omega)-\frac{1}{2}e^{\hbar \beta \omega}\ft{\Delta}_{i j}(\omega) 
% \end{align}
% which reduces to:
\begin{equation}
    \ft{G}^S_{i j}(\omega)=\frac{\hbar}{2}\coth{\frac{\hbar \beta \omega}{2}}\ft{\Delta}_{i j}(\omega),
\end{equation}
which, in terms of the retarded Green's function, reduces to the familiar quantum FDT:
\begin{align}\label{quantumFDT}
    \ft{G}^S_{i j}(\omega)
    &=\hbar\coth{\frac{\hbar \beta \omega}{2}}\mathrm{Im} \ft{G}^R_{i j}(\omega) 
\end{align}
since
\begin{equation}
   \mathrm{Im} \ft{G}^R(\omega)  = \frac{\ft{\Delta}(\omega)}{2\hbar}.
\end{equation}
Note that the factor 2 dividing is due to the retarded Green's function defined only for positive time, see \ref{FT}. 

Finally, in the classical or high temperature limit, we obtain the classical FDT:
\begin{align}\label{eq:classicalFDT}
    \ft{G}^S_{i j}(\omega)= \frac{2}{\beta \omega} \mathrm{Im} \ft{G}^R_{i j}(\omega), \quad     
    (\hbar \rightarrow 0 \; \text{and/or} \; \beta \rightarrow 0 ).
\end{align}

% Back to the time space, in the case when $\mathrm{Im} \ft{G}^R_{i j}(-\omega)=-\mathrm{Im} \ft{G}^R_{i j}(\omega)$, then:
% \begin{align}\label{FDT}
%     G^S_{ij}(t) &= -\frac{1}{2 \pi} \int_{-\infty}^{\infty} d\omega \,  \hbar\coth{\frac{\hbar \beta \omega}{2}}\mathrm{Im} \ft{G}^R_{i j}(\omega) e^{i \omega t}\\
%     &=-\frac{\hbar}{\pi} \int_{0}^{\infty} d\omega \, \coth{\frac{\hbar \beta \omega}{2}}\mathrm{Im} \ft{G}^R_{i j}(\omega) \cos( \omega t) 
% \end{align}

%% file: Appendix/GLE.tex
\section{Generalized Langevin equation}\label{sec:Langevin}
A quantum or classical harmonic oscillator coupled linearly to a harmonic bath is an exactly solvable model \cite{fleming2011exact, ford1988quantum}, and, as discussed in section \ref{sec:QBM}, can be described in terms the generalized Langevin equation:
\begin{align}\label{quantum_Langevin_with_dissipation_kernel}
&M \ddot{q}(t)+M (\Omega^{2}+ \delta \Omega^{2} ) q(t)+\int_{0}^{t} d \tau \mu(t-\tau) q(\tau) = \noise(t)
\end{align}
with the initial conditions:
\begin{equation}
    q(0)=q_{0}, \quad \dot{q}(0)=\dot{q}_{0}.
\end{equation}
and a Gaussian noise:
\begin{align}
     \langle \noise(t) \rangle &=0\\
     \frac{1}{2}\braket{\{\noise(t), \noise(t')\}}&=\nu(t-t').
\end{align}

% Notice that, if $I(\omega)=-I(-\omega)$, then the above relationship leads to
% \begin{align}
%     \gamma(t)
%     &=\frac{1}{2 M} \int_{0}^{\infty} d \omega  \frac{I(\omega)}{\omega} (e^{i \omega t}+e^{-i \omega t})\\
%     &=\frac{1}{2 \pi} \int_{-\infty}^{\infty} d\omega  \frac{\pi I(\omega)}{M \omega}  e^{i \omega t},
% \end{align}
% i.e. the Fourier transform of the damping kernel is related to the spectral density:
% \begin{equation} \label{damping_kernel_spectral_density}
%     \boxed{\tilde{\gamma}(\omega) = \frac{\pi I(\omega)}{M \omega}} .
% \end{equation}

% \subsubsection{Derivation of the Langevin equation}
% The non-local term of the Langevin equation \eqref{quantum_Langevin_with_dissipation_kernel}:
% \begin{align}
%   2 \int_{0}^{t} d \tau \mu(t-\tau) x(\tau) 
%     &=2 M \int_{0}^{t} d \tau  \frac{\partial \gamma(t-\tau)}{\partial t}  x(\tau)\\
%     &= -2M \int_{0}^{t} d \tau  \frac{\partial \gamma(t-\tau)}{\partial \tau}  x(\tau)\\
%     &= -2M  \left.\left(\gamma(t-\tau) x(\tau)\right) \right|^t_0
%     +  2M \int_{0}^{t} \gamma(t-\tau) \dot{x}(\tau)\\
%     &= -2M  \left(\gamma(0) x(t)-\gamma(t) x(0)\right) 
%     + 2 M \int_{0}^{t} \gamma(t-\tau) \dot{x}(\tau).
% \end{align}

% The boundary term is:
% \begin{equation}
%     2 M \gamma(0)=2 \int_{0}^{\infty} d \omega  \frac{I(\omega)}{\omega}=d\Omega^2.
% \end{equation}

\subsection{Solution}
It is common in the literature to find the generalized Langevin equation in terms of the damping kernel $\gamma$, aka the memory function, which is defined as:
\begin{equation}
     \mu(t-\tau)=M \frac{\partial}{\partial t} \gamma(t-\tau)
\end{equation}
then:
\begin{equation}\label{quantum_Langevin_with_damping_kernel}
\ddot{q}(t)+\Omega^{2} q(t)+\int_{0}^{t} d \tau \gamma(t-\tau) \dot{q}(\tau)+\gamma(t) q(0)=\noise(t)/M
\end{equation}

The generalized Langevin equation \eqref{quantum_Langevin_with_damping_kernel} can be solved using Laplace transform, which gives:
\begin{equation}
    (s^{2}+\Omega^{2}+s \hat{\gamma}(s)) \hat{q}(s)=s q_{0}+\dot{q}_{0}
    +\hat{\noise}(s)/M
\end{equation}
from which the solution is:
\begin{equation}
     \hat{q}(s)=M (s \,q_{0}+\dot{q}_{0})\hat{G}(s)
    +\hat{\noise}(s) \hat{G}(s)
\end{equation}
where the Laplace transform of the (retarded) Green's function is
% \begin{align}
%     \lt{G}^R(s)
%     &=\frac{1/M}{s^{2}+\Omega_R^{2}+ \lt{\mu}(s)/M}
% \end{align}
\begin{equation}\label{Greensfunction_Laplace}
   \hat{G}(s)= \dfrac{1/M}{s^{2}+\Omega^{2}+s \hat{\gamma}(s)},
\end{equation}
which, in time space, must satisfy the initial conditions
\begin{equation}\label{Greensfunction_initial_conditions}
    G(0)=0, \quad \dot{G}(0)=\frac{1}{M}
\end{equation}
that fully determine it.

The solution back to the time space is:
\begin{equation} \label{solution_langevin}
q(t)=M\left(q_{0} \dot{G}(t)+\dot{q}_{0} G(t)\right)+\int_{0}^{t} d \tau G(t-\tau) \noise(\tau).
\end{equation}

\subsection{Stationary limit}
The general solution \eqref{solution_langevin} simplifies in the stationary limit, reducing to:
\begin{align}\label{stationary_solution}
    q_s (t) = \int_{0}^{\infty} d \tau G(t-\tau) \noise(\tau), \quad (t \rightarrow \infty),
\end{align}
that means for non-vanishing initial conditions, we require $G(t\longrightarrow \infty)=0$ and $\dot{G}(t\longrightarrow \infty)=0$. However, if $q_0=0$ then $G(t\longrightarrow \infty)=0$ is sufficient to reach stationarity. 

The causality condition of the Green's function \eqref{eq:causalityG} allows the integration to be extended to the full real axis. Hence, upon Fourier transform, the stationary solution \eqref{stationary_solution} is just a product:
\begin{equation}\label{eq:FT_stationary}
    \ft{q}_s(\omega) = \ft{G}(\omega) \ft{\noise}(\omega).
\end{equation}
% where 
% \begin{equation}
%    \ft{G}(\omega)
%    =\frac{1/M}{-\omega^{2}+\Omega^{2}+\ft{\mu}(\omega)/M}
% \end{equation} 
% which solves the stationary generalized Langevin equation:
% \begin{align}
% &M \ddot{q}(t)+M \Omega^{2} q(t)+\int_{-\infty}^{\infty} d \tau \mu(t-\tau) q(\tau) = \noise(t).
% \end{align}

The gFDR \eqref{eq:GsImGr_QBM} can be readily obtained using the definition of the correlation function and the stationary solution \eqref{eq:FT_stationary}.

%% file: Appendix/notationConvention.tex
\section{Fourier transform}\label{FT}
Consider the following definitions of Fourier transform:
\begin{align}
    \ft{f}(\omega)
    % &\equiv\mathrm{FT}\{f(t)\}(\omega)
    &\equiv \int_{-\infty}^\infty dt \,e^{-i \omega t} f(t)\\
    % \ft{f}_{\pm}(\omega)
    % &\equiv \int_{-\infty}^\infty dt \,e^{-i \omega t} f(t)\theta(\pm t),
    \ft{f}_{+}(\omega)
    &\equiv \int_{0}^\infty dt \,e^{-i \omega t} f(t)\\
    \ft{f}_-(\omega)
    &\equiv \int^{0}_{-\infty} dt \,e^{-i \omega t} f(t)
\end{align}
such that
\begin{align}
    \ft{f}(\omega)
    &=\ft{f}_-(\omega)+\ft{f}_+(\omega).
\end{align}
It follows that:
\begin{enumerate}
    \item For functions even in time $f(t)=f(-t)$:
    \begin{equation}
        \ft{f}_{-}(\omega)=\ft{f}_{+}(-\omega)\Longrightarrow \ft{f}(\omega)=\ft{f}(-\omega)
    \end{equation}
    \item For functions odd in time $f(t)=-f(-t)$:
    \begin{equation}
        \ft{f}_{-}(\omega)=-\ft{f}_{+}(-\omega)
        \Longrightarrow \ft{f}(\omega)=-\ft{f}(-\omega)
    \end{equation}
    
    \item If $\ft{f}_-(\omega)=\ft{f}_+(\omega)$,
then $\ft{f}(\omega)=2\ft{f}_+(\omega)$.
    \item If $g(t)=f(t)\theta(t)$, then $\ft{g}(\omega)=\ft{f}_{+}(\omega)$.
\end{enumerate}

% \subsection{Laplace transform}

% The Laplace transform:
% \begin{equation}
%     \lt{f}(s)=\mathcal{L}_s\{f(t)\}=\int_{0}^{\infty} d t e^{-s t} f(t)
% \end{equation}

% The convolution theorem:
% \begin{align}
%     \mathcal{L}_s \left\{ \int_0^t dt' \,f(t-t') g(t')\right\} = \lt{f}(s)\lt{g}(s)
% \end{align}
% The convolution with identity gives:
% \begin{align*}
%     \mathcal{L}_s \left\{ \int_0^t dt' \,f(t')\right\} = \lt{f}(s)/s
% \end{align*}